\documentstyle[preprint,aps,floats,psfig]{revtex}
\begin{document}
\newenvironment{tab}[1]
{\begin{tabular}{|#1|}\hline}
{\hline\end{tabular}}

\def\sfhv{\hat{\mbox{\boldmath${\scriptstyle p}$}}_{\!\!{\scriptscriptstyle F}}}
\def\pfhv{\hat{\mbox{\boldmath$p$}}_{\!\!{\scriptscriptstyle F}}}
\newcommand{\base}{}
\newcommand{\nonb}{\nonumber}
\newcommand{\RA}{\rangle}
\newcommand{\LA}{\langle}
\newcommand{\LL}{\langle \langle}
\newcommand{\RR}{\rangle \rangle}
\newcommand{\HG}{\hat{G}}
\newcommand{\HK}{\hat{K}}
\newcommand{\Ht}{\hat{t}}
\newcommand{\HA}{\hat{A}}
\newcommand{\HI}{\hat{I}}
\newcommand{\HX}{\hat{X}}
\newcommand{\HY}{\hat{Y}}
\newcommand{\Hg}{\hat{g}}
\newcommand{\Hrho}{\hat{\rho}}
\newcommand{\DA}{{\cal D}^A}
\newcommand{\DR}{{\cal D}^R}
\newcommand{\CG}{{\cal G}}
\newcommand{\D}{{\cal D}}
\newcommand{\W}{{{\cal W}_{n,m}(N_a,N_b)}}
\newcommand{\vk}{{\vec{k}}}
\newcommand{\vx}{{\vec{x}}}

\title  {Transport properties of ferromagnet-$d$-wave 
superconductor ferromagnet double junctions}
\author{N. Stefanakis and R. M\'elin}
\address{Centre de Recherches sur les Tr\`es Basses Temp\'eratures, 
Centre National de la Recherche Scientifique,
25 Avenue des Martyrs, BP 166, 38042 Grenoble c\'{e}dex 9, France}
\date{\today}
\maketitle

\begin{abstract}
We investigate transport properties of a trilayer made of
a $d$-wave superconductor connected to two ferromagnetic electrodes.
Using Keldysh formalism we show that crossed Andreev 
reflection and elastic cotunneling 
exist also with $d$-wave superconductors. Their 
properties are controlled by the existence of zero energy states
due to the anisotropy of the $d$-wave 
pair potential. 

\end{abstract}
\section{introduction} 
Transport in superconductor-ferromagnetic hybrid systems 
has received great attention in the last years due to the 
progress in nanotechnology which 
made possible the fabrication and 
characterization of various heterostructures 
\cite{soulen,upadhyay,giroud,moussy,bourgeois}. 

During the last years phase sensitive tests have shown 
that the order parameter in cuprate superconductors is 
predominantly of a $d$-wave 
symmetry \cite{kirtley,tsuei,hilgenkamp,kashiwayatanaka}. 
In $d$-wave superconductors the zero bias 
conductance peak (ZBCP) observed in the tunneling spectra 
results from zero energy states (ZES) 
that are formed due to the sign change of the order 
parameter in orthogonal directions in $k$ space. The ZBCP depends 
on the orientation of the surface and does not exist for 
$s$-wave superconductors
\cite{fogelstrom,covington,aprili,krupke,stefan1,stefan2}.

In $d$-wave superconductor ferromagnet junctions the ZBCP is 
suppressed by the increase of the exchange field of the ferromagnet
\cite{stefan,zhu,kashiwaya,zutic}. 
This can be understood from the fact that increasing spin 
polarization in the ferromagnetic electrode suppresses
Andreev reflection and therefore suppresses the ZBCP
which is due to the fact that the transmitted quasiparticles
are subject to the sign change of the order parameter.
Moreover  
in ferromagnet/$d$-wave superconductor/ferromagnet double junctions, 
the quasiparticle current is enhanced compared to the normal state
because of ZES
\cite{yoshida}. Very recently a circuit theory of 
unconventional superconductors has been presented \cite{tanaka}.

Keldysh formalism has been 
applied to normal metal-$s$-wave superconductor
junctions \cite{cuevas}, and 
in multiterminal configurations where one $s$-wave superconductor is 
connected to several ferromagnetic electrodes \cite{melin2,melin1,deutscher,falci}. 
The conductance of multiterminal hybrid structures is due to two
types of processes: (i) crossed Andreev reflection in which Cooper
pairs are extracted from the superconductor. The spin-up electron
of the Cooper pair tunnels in a spin-up ferromagnet and the spin-down
electron tunnels in a spin-down ferromagnet; (ii) elastic cotunneling
in which a spin $-\sigma$ electron from one electrode is transferred
as a spin-$\sigma$ electron in another electrode.

The purpose of the present work is to investigate transport 
properties of a 

ferromagnet-$d$-wave superconductor-ferromagnet 
double junction via Keldysh formalism.
We find that crossed Andreev reflection and elastic cotunneling
are influenced by the $d$-wave symmetry of the order parameter
in the sense that both processes are mediated by zero energy
states formed for certain orientation of the $d$-wave order
parameter.

The article is organized as follows. In Sec. II 
we introduce surface Green's functions. In Secs. III, IV 
we describe transport theory and present the results. 
Concluding remarks are given in the last section.

\section{surface Green's functions}

The quasiparticle properties of $d$-wave superconductors 
are influenced by interfaces and surfaces: due to 
the anisotropy of the order parameter
the quasiparticles that are reflected from the surface or 
transmitted through the interface are subject to the sign change 
of the order parameter. Therefore surface properties are 
different from bulk properties and we use in transport theory
the surface Green's functions that take into account 
the contributions of all waves that propagate close to the
surface \cite{samanta,matsumoto}. Green's function techniques 
have been used to calculate the conductance of $d$-wave 
superconductors near impurities \cite{bayers,salkola}.
Also surface quasiclassical Green's functions have been used in the 
calculation of the Josephson current between $d$-wave superconductors
\cite{barash,barash2,cuevas2,fominov}. 
In our case this Green's function is inserted in
a three node balistic circuit
that is used to describe the transport
properties of the hybrid structure containing a $d$-wave superconductor.

The form of the local retarded surface Green's function matrix for 
a $d$-wave superconductor is the following, 
for smooth interface, with momentum conservation in the plane of the 
interface
\begin{equation}
\hat g^{xx,R}(E,\theta)=
\left(\begin{array}{cc}
g & f\\
\bar{f} & g
\end{array}\right).
\label{dwaveg}
\end{equation}
It obeys the Eilenberger equation \cite{eilenberger} and satisfies the 
normalization condition $\hat g^2=1$. It can be parametrized 
as follows 
\begin{equation}
g=\frac{1-ab}{1+ab}, f=\frac{2a}{1+ab}, \bar{f}=\frac{2b}{1+ab},
\label{schopohl}
\end{equation}
where the $a(x,\theta)$ and $b(x,\theta)$ satisfy the Riccati 
equations \cite{schopohlmaki} 
\begin{equation}
\hbar v_F \cos (\theta) \frac{da}{dx}-2iEa+\Delta^{*}a^2-\Delta=0
\end{equation}
\begin{equation}
\hbar v_F \cos (\theta) \frac{db}{dx}+2iEb-\Delta b^2+{\Delta}^{*}=0,
\end{equation} 
where $v_F$ is the Fermi velocity. We assume for simplicity 
that the gap function $\Delta$ is constant. Then the spatially 
independent $a,b$ functions are found as 
\begin{equation}
a(0,\theta)=\frac{i(E-\epsilon_{+} sgn E)}{\Delta_{+}(\theta)}
\end{equation}
\begin{equation}
b(0,\theta)=\frac{i(E-\epsilon_{-} sgn E)}{\Delta_{-}^{*}(\theta)},
\end{equation}
where $\theta$ is the angle between the normal to the interface
and the trajectory of the quasiparticle.
$\Delta_{+}(\theta)=\Delta(\theta)$
($\Delta_{\_}(\theta)=\Delta(\pi- \theta)$) is the
pair potential experienced by the
quasiparticle along the trajectory $\theta (\pi-\theta)$ and
$\epsilon_{\pm}=\sqrt{E^2-\Delta_{\pm}(\theta)^2}$. 
In case of $d_{x^2-y^2}$-wave superconductor
\begin{equation}
\Delta(\theta)=\Delta_0
\cos[2(\theta - \beta)]
,~~~\label{deltad}
\end{equation}
where $\beta$ denotes the angle between the normal to the interface
and the $x$-axis of the crystal. Then $g$ and $f$ Green functions 
are calculated from Eqs. \ref{schopohl}.

\begin{figure}
\begin{center}
\centerline{
\psfig{figure=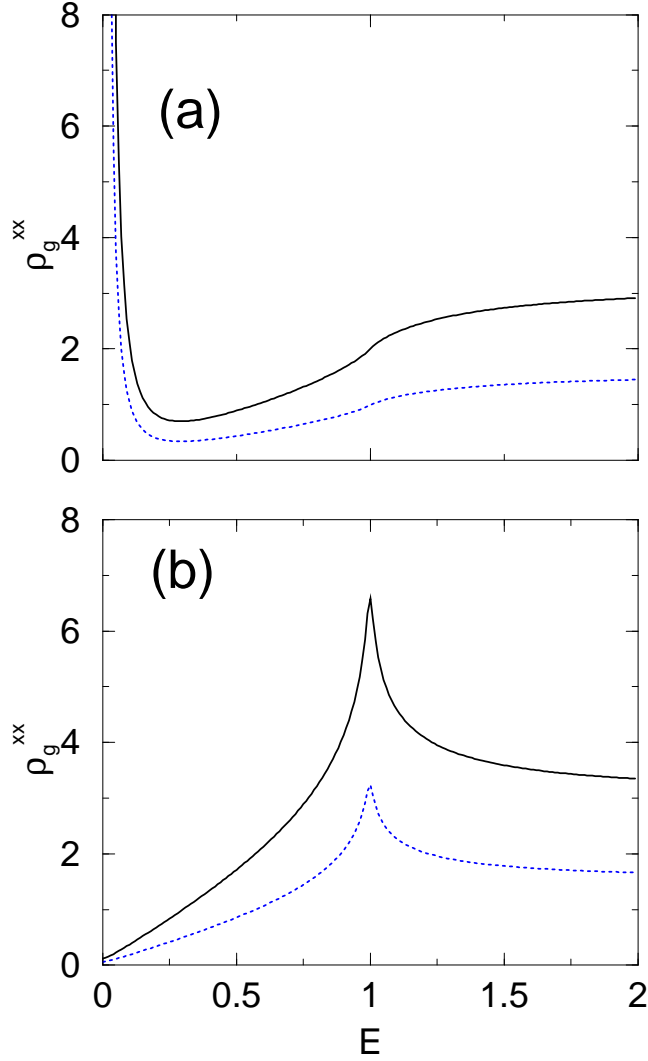,width=8.5cm}}
\end{center}
\caption{(a) The solid line represents averaged over the Fermi surface 
density of states $\rho_g^{xx}$ for $\beta=\pi/4$. 
A well defined ZEP exists.
We have put a small imaginary part $\delta=0.01$ in the 
energy $\omega$ of the Green's function. 
The dashed line represents the $\rho_g^{xx}$ averaged 
over the Fermi surface with 
an angular depended transmission coefficient. 
(b) The same as in (a) but for $\beta=0$
}
\label{dos.fig}
\end{figure}

In fact in the transport equations
we should average over the Fermi surface in order to include the 
details of the order parameter symmetry. 
We have calculated the density of states averaged over the Fermi surface
$\rho_g^{xx}=<Re g^{xx,R}(E,\theta,)>$. 
In Fig. \ref{dos.fig} the density of states $\rho_g^{xx}$ is plotted for 
different orientations of the order parameter $\beta=0$,
and $\beta=\pi/4$.
A ZEP is formed for $\beta=\pi /4$ due to sign change of the 
pair potential. A small imaginary or effectively dissipative term ($\delta=0.01$) was added in the 
energy in order to make this peak visible. 

In order to study the effect of the angular dependence of 
the transmission coefficient we calculate the Fermi surface 
averaged density of states $\rho_g^{xx,D}$ defined as
$$
\rho_g^{xx,D} = \int_{-\pi/2}^{\pi/2}
d \theta D(\theta) (Re g^{xx,R}(E,\theta))
,
$$
where $D(\theta)=\sin^2(\theta)$ \cite{luck}. 
We see that the 
density of states is suppressed when this coefficient is included in the 
calculation (see Fig.~\ref{dos.fig}). However we do not expect the transport properties to change 
qualitatively compared to the case where the transmission is 
independent on the angle (see Fig.~\ref{dos.fig}). In the following 
we use $D(\theta)=1$.
Transport can probe the symmetry of the $d$-wave order parameter 
if we consider the orientation of the $d$-wave 
order parameter $\beta$ as a variable.

The ferromagnetic electrodes are described by the Green's function
\begin{equation}
\label{eq:single-Ferro}
\Hg^{R,A} = \mp i \pi
\left[ \begin{array}{cc}
\rho_{1,1} & 0 \\
0 & \rho_{2,2}\end{array} \right]
,
\end{equation}
where $\rho_{1,1}$ and $\rho_{2,2}$
are respectively the spin-up and
spin-down densities of states.

\section{transport theory}
\begin{figure}
\begin{center}
\centerline{
\psfig{figure=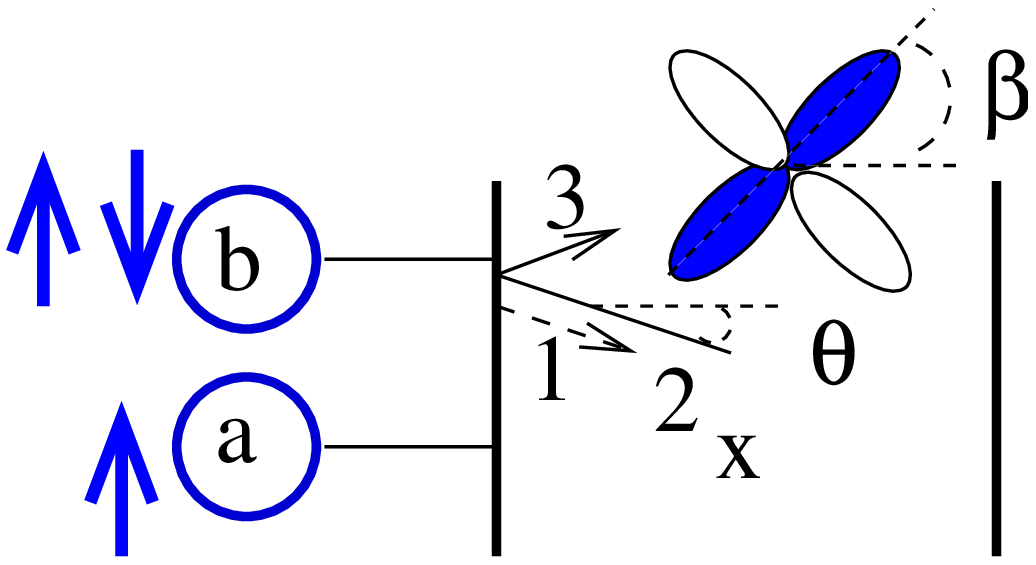,width=7.5cm,angle=0}}
\end{center}
\caption{
The geometry of the junction involving the d-wave superconductor
and two ferromagnetic electrodes of full spin polarization
as indicated in the figure by up and down arrows. 
The orientation of the magnetization 
of the ferromagnetic electrodes can be parallel or antiparallel. 
The excitation at $x$ gives rise to several outgoing trajectories $\theta$. 
In the figure only one of these trajectories is presented. $\beta$ is the 
orientation of the $d$-wave order parameter shown also in the figure 
with respect to the 
direction $x$. The electron like quasiparticle $2$ is reflected as 
a hole like quasiparticle $1$ and an electron like quasiparticle $3$.
The labels $a,b$ in the figure correspond to the electrods 
$a,b$ respectively.
}
\label{circuit.fig}
\end{figure}

We use a Green's functions method to describe transport 
in a system made of two ferromagnetic electrodes connected to a 
$d$-wave superconductor (see Fig. \ref{circuit.fig}).
We first solve the Dyson equation which in a 
$2 \times 2$ Nambu representation has the following form 
for the advanced ($\HG^A$) and retarded ($\HG^R$)  
Green's functions~\cite{keldysh,caroli}

\begin{equation}
\label{eq:Dy1}
\HG^{R,A} = \Hg^{R,A} + \Hg^{R,A} \otimes \hat{\Sigma}
\otimes \HG^{R,A}
.
\end{equation}
$\hat{\Sigma}$
is the self energy that contains the coupling of the tunnel
Hamiltonian. The tunnel Hamiltonian associated to
Fig.~\ref{circuit.fig} takes the form
$$
{\cal W} = \sum_\sigma \left[ t_{a,x} c_a^+ c_x
+t_{x,a} c_x^+ c_a + t_{b,x} c_b^+ c_x
+t_{x,b} c_x^+ c_b \right]
.
$$
$\Hg$ in Eq.~(\ref{eq:Dy1}) is the Green's functions
of the disconnected system ({\sl i.e.,} with $\hat{\Sigma}=0$).
The symbol $\otimes$ includes a summation over the nodes of
the network and a convolution over time arguments. Since we
consider stationary transport this conclusion is transformed
into a product by Fourier transform.
$\HG$ is
the Green's functions of the connected system
({\sl i.e.,} with $\hat{\Sigma} \ne 0$).
The Keldysh component is given by~\cite{caroli}
\begin{equation}
\label{eq:Dy2}
\HG^{+,-} = \left[ \HI + \HG^R \otimes \hat{\Sigma}
\right] \otimes \Hg^{+,-} \otimes
\left[ \HI + \hat{\Sigma} \otimes
\HG^{A} \right]
.
\end{equation}
The current is related to the
Keldysh Green's function~\cite{caroli} by the relation
\begin{equation}
\label{eq:I-gene}
I_{a,x} = \frac{e}{h}
\int d \omega \left[ \Ht_{a,x} \HG^{+,-}_{x,a}
- \Ht_{x,a} \HG^{+,-}_{a,x} \right]\sigma^z
.
\end{equation}
At this stage of the
calculation no explicit angular form of the 
tunneling matrix elements was assumed. The effect of the 
angular dependence of the transmission coefficient in the transport 
properties was already discussed in the previous section.
The elements of the 
differential conductance matrix that we want to calculate
are given by
\begin{equation}
\label{eq:def-mat-elem}
\CG_{a_i,a_j}(V_a,V_b) =
\frac{ \partial I_{a_i}}{\partial V_{a_j}}
(V_a,V_b)
.
\end{equation}
The principle of the calculation of $\CG_{a_i,a_j}(V_a,V_b)$ is 
similar to the $s$-wave case \cite{melin2}.
Depending on the orientation of the magnetizations in the 
two ferromagnetic electrodes we can distinguish the 
following cases:

\subsection{Antiparallel magnetizations}
If the two ferromagnetic
electrodes have an antiparallel spin orientation we find
for the elements of the conductance matrix
\begin{eqnarray}
\label{Gf}
\label{f-qp}
{\cal G}_{a,a} &=& + 4 \pi^2 |t_{a,x}|^2
\rho^{a,a}_{1,1} \rho_g^{x,x}\\\nonb
&&\times
\frac{1}{\DA \DR}
\left[ 1 - |t_{b,x}|^2
g^{b,b,A}_{2,2}
g^{x,x,A} \right]
\left[ 1 - |t_{b,x}|^2
g^{b,b,R}_{2,2}
g^{x,x,R} \right]\\
\label{f-mixed1}
&-& 2 i \pi |t_{a,x}|^2
|t_{b,x}|^2
\rho^{a,a}_{1,1}g_{2,2}^{b,b,A}
\\\nonb&&\times \frac{1}{\DA \DR}
f^{x,x,A}
f^{x,x,A}
\left[ 1 - |t_{b,x}|^2
g^{b,b,R}_{2,2}
g^{x,x,R} \right]\\
\label{f-mixed2}
&+& 2 i \pi |t_{a,x}|^2
|t_{b,x}|^2
\rho^{a,a}_{1,1}
g_{2,2}^{b,b,R} \\\nonb&&\times \frac{1}{\DA \DR}
 f^{x,x,R}
f^{x,x,R}
\left[ 1 - |t_{b,x}|^2
g^{b,b,A}_{2,2}
g^{x,x,A} \right]
,
\label{f-andreev}
\end{eqnarray}
and
\begin{eqnarray}
{\cal G}_{a,b} &=& -4 \pi^2 |t_{a,x}|^2
|t_{b,x}|^2
\frac{1}{\DA \DR}
\rho_{1,1}^{a,a}
\rho_{2,2}^{b,b}
f^{x,x,R} f^{x,x,A}
.\label{careGab}
\end{eqnarray}
The expression of the determinant
${\cal D}_{\rm R}$ is the following:
\begin{equation}
\label{Df}
{\cal D}_{\rm R} =
1 - |t_{b,x}|^2
g^{b,b,R}_{2,2}
g^{x,x,R}-
|t_{a,x}|^2
g^{a,a,R}_{1,1}
g^{x,x,R}+
|t_{b,x}|^2 |t_{a,x}|^2
g^{a,a,R}_{1,1}
g^{b,b,R}_{2,2}
({g^{x,x,R}}^2-{f^{x,x,R}}^2), 
\end{equation}
and a similar expression holds for ${\cal D}_{\rm A}$. 
$\rho^{a,a}_{1,1}, \rho^{b,b}_{2,2}, \rho_g^{x,x}$ are the density of 
states of electrodes $a$,$b$ and the superconductor respectively. 
Contrary to the $s$-wave case $\rho_g^{x,x}$ is not 
zero for $E<\Delta_0$ and the term (\ref{f-qp}) contributes
also to the 
quasiparticle current even below the superconducting gap. 
In the $s$-wave case there are simple relations between the 
conductance matrix elements 
(for instance ${\cal G}_{a,a}={\cal G}_{a,b}$) which means that the 
transport is mediated only by Cooper pairs \cite{melin2}. In the $d$-wave case 
such relations are no more valid because of the quasiparticle tunneling.
Also depending on the trajectory angle $\theta$ 
the different terms contribute to the Andreev current and the 
quasiparticle current. Moreover in the present case the propagators 
$f^{x,x,A,(R)},g^{x,x,A(R)}$ have a $d$-wave symmetry.  
In (\ref{careGab}) $G_{ab}$ depends on $f^{x,x,R} f^{x,x,A}$ and therefore
the corresponding matrix element is associated to 
crossed Andreev reflections. 

\subsection{Parallel magnetizations}
If the electrodes
have a parallel spin orientation, we find
\begin{eqnarray}
\label{Gg}
\label{g-qp}
{\cal G}_{a,a} &=& + 4 \pi^2 |t_{a,x}|^2
\rho^{a,a}_{1,1} \rho_g^{x,x}\\\nonb
&&\times
\frac{1}{\DA \DR}
\left[ 1 - |t_{b,x}|^2
g^{b,b,A}_{1,1}
g^{x,x,A} \right]
\left[ 1 - |t_{b,x}|^2
g^{b,b,R}_{1,1}
g^{x,x,R} \right]\\
\label{g-mixed1}
&-& 2 i \pi |t_{a,x}|^2
|t_{b,x}|^2
\rho^{a,a}_{1,1}g_{1,1}^{b,b,A}
\\\nonb&&\times \frac{1}{\DA \DR}
g^{x,x,A}
g^{x,x,A}
\left[ 1 - |t_{b,x}|^2
g^{b,b,R}_{1,1}
g^{x,x,R} \right]\\
\label{g-mixed2}
&+& 2 i \pi |t_{a,x}|^2
|t_{b,x}|^2
\rho^{a,a}_{1,1}
g_{1,1}^{b,b,R} \\\nonb&&\times \frac{1}{\DA \DR}
g^{x,x,R}
g^{x,x,R}
\left[ 1 - |t_{b,x}|^2
g^{b,b,A}_{1,1}
g^{x,x,A} \right],
\label{g-andreev}
\end{eqnarray}
and
\begin{eqnarray}
{\cal G}_{a,b} &=& -4 \pi^2 |t_{a,x}|^2
|t_{b,x}|^2
\frac{1}{\DA \DR}
\rho_{1,1}^{a,a}
\rho_{1,1}^{b,b}
g^{x,x,R} g^{x,x,A}
.\label{cotuGab}
\end{eqnarray}
The determinant ${\cal D}_{\rm R}$ is given by
\begin{equation}
\label{Dg}
{\cal D}_{\rm R} =
1 - |t_{b,x}|^2
g^{b,b,R}_{2,2}
g^{x,x,R}-
|t_{a,x}|^2
g^{a,a,R}_{1,1}
g^{x,x,R}.
\end{equation}

In (\ref{cotuGab}) $G_{ab}$ depends on $g^{x,x,R} g^{x,x,A}$ and therefore
the corresponding matrix element is associated to 
cotunneling processes.
The elements ${\cal G}_{b,a},{\cal G}_{b,b}$ of the 
conductance matrix which describe transport through electrode 
$b$ 
are derived from the corresponding expressions 
for ${\cal G}_{a,a},{\cal G}_{a,b}$ by the substitution $a\leftrightarrow b$
for the parallel alignment. For the antiparallel alignment the 
following set of substitutions should be made:
${ g^{b,b,A(R)}_{2,2} \leftrightarrow g^{a,a,A(R)}_{1,1}, 
t_{a,x} \leftrightarrow t_{b,x}, \mu_a \leftrightarrow \mu_b }$.

\section{Results}

\subsection{Antiparallel magnetizations}
\begin{figure}
\begin{center}
\leavevmode
\centerline{
\psfig{figure=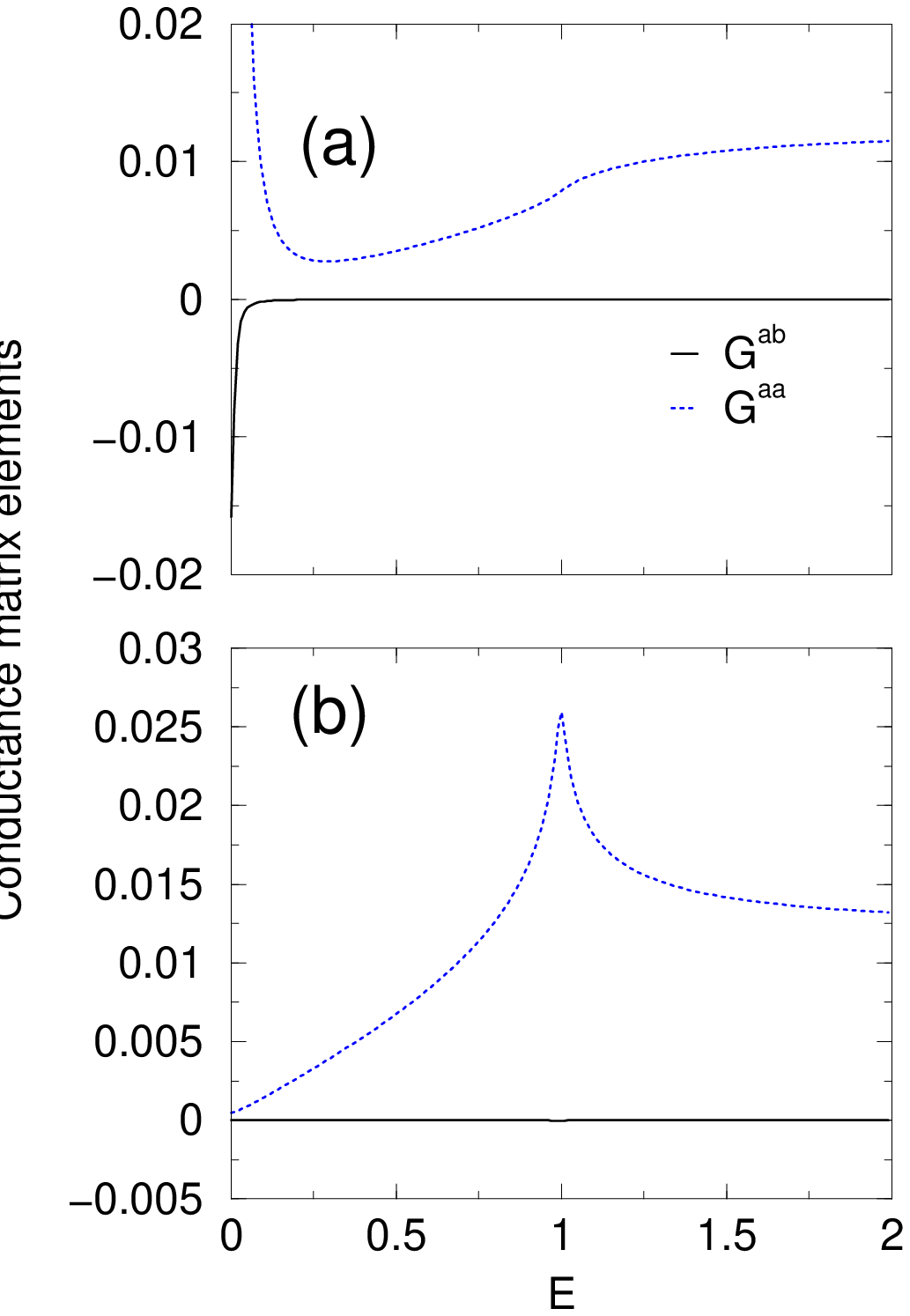,width=8.5cm}
}
\end{center}
\caption{
Conductance ${\cal G}^{ab},{\cal G}^{aa}$ for the antiparallel spin 
orientation of the ferromagnetic electrodes,
as a function of $E$ (in units of $\Delta_0$)
for different orientations of the $d$-wave order parameter
(a) $\beta=\pi/4$, (b) $\beta=0$.
The hopping element is $0.01$.
}
\label{care.fig}
\end{figure}

We consider the ferromagnet/$d$-wave superconductor/ferromagnet 
double junction shown in Fig. \ref{circuit.fig}. 
For the antiparallel alignment of the 
magnetizations in the two ferromagnetic electrodes the conductance 
depends on the orientation $\beta$ as well as on the transparencies of the 
interfaces $t_{a,x}$, $t_{b,x}$. 
For $\beta=\pi/4$ (see Fig. \ref{care.fig}(a)) the surface Green's function 
has a pole at $E=0$ and the conductance (both ${\cal G}^{aa}$ and 
${\cal G}^{ab}$) 
acquires a ZEP.
The conductance ${\cal G}^{aa}$ above the gap 
depends only on the density of states 
$\rho_g^{xx}$ and for large energies it has a finite value.
${\cal G}^{ab}$ 
depends only on crossed Andreev reflection processes and 
is zero above the gap. 

For $\beta=0$ (see Fig. \ref{care.fig}(b))
similarly to the $s$-wave case no ZES are formed at the 
interface and both ${\cal G}^{aa}$ and ${\cal G}^{ab}$ 
take relatively small values. 
However the line shape of the conductance is $V$ and is determined by 
$\rho_g^{xx}$. 
In the $s$-wave case the line shape of the conductance is $U$ and 
a peak just below the energy gap exists \cite{melin2}. 
In this sense the results 
for $s$-wave are qualitatively different than 
for $d$-wave with $\beta=0$
due to the anisotropy of the $d$-wave order parameter.

To summarize
transport for antiparallel magnetizations is due 
to crossed Andreev reflection in which a spin-up electron 
from one electrode is transferred as a spin-down
hole in the 
other electrode, and is influenced by
ZES that are formed at the interface due to the 
sign change of the order parameter. The enhancement of the 
quasiparticle current at $E=0$ for $\beta=\pi/4$ in the
$d$-wave superconductor ferromagnet double junction has also been 
found recently using the scattering approach \cite{yoshida}.

\subsection{Parallel magnetizations}
\begin{figure}
\begin{center}
\leavevmode
\centerline{
\psfig{figure=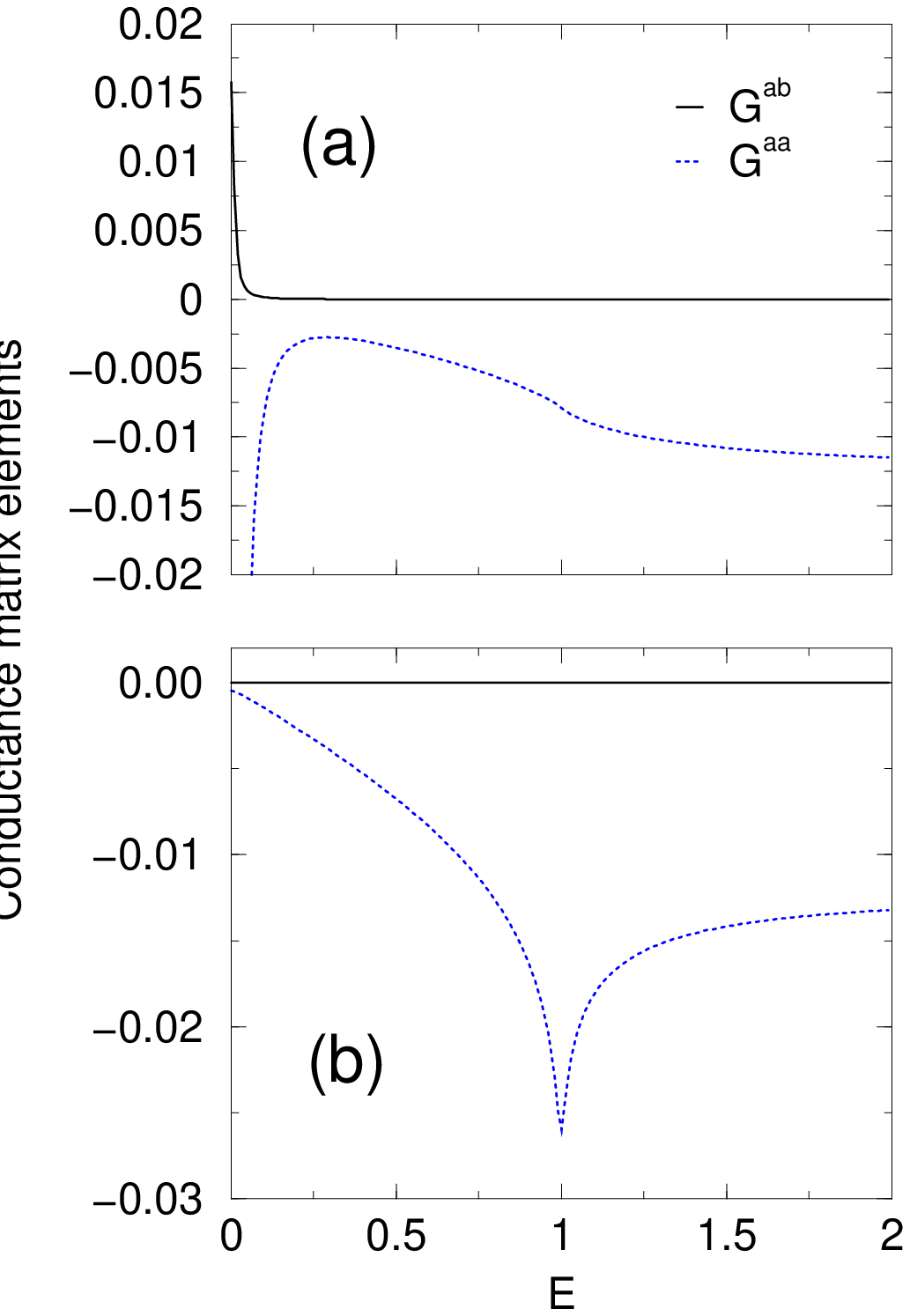,width=8.5cm}
}
\end{center}
\caption{
Conductance ${\cal G}^{ab},{\cal G}^{aa}$ for the parallel spin
orientation of the ferromagnetic electrodes,
as a function of $E$ (in units of $\Delta_0$)
for different orientations of the $d$-wave order parameter
(a) $\beta=\pi/4$, (b) $\beta=0$.
The hopping element is $0.01$.
}
\label{cotu.fig}
\end{figure}

The results concerning the ZES are not modified qualitatively when the 
orientation of the magnetizations is parallel (see Fig. \ref{circuit.fig}). 
The conductance above the gap is determined mainly by $\rho_g^{xx}$. 
For $\beta=0$ (see Fig. \ref{cotu.fig}(b)) 
the results are similar to the case of the 
antiparallel alignment.  

To summarize transport for parallel magnetizations
is due to elastic
cotunneling in which an electron
from electrode $a$ is transmitted as an electron in electrode $b$, 
and is influenced by
the ZES that are formed for certain orientation of the $d$-wave order 
parameter. 
For $\beta=\pi/4$ the interface at large values of the barrier strength
becomes transparent due to bound states formed because of the sign 
change of the order parameter in orthogonal directions in $k$-space.
This property does not exist for $s$-wave superconductors.

\section{relevance to experiment}
Multiterminal superconductor ferromagnet structures 
can be used to test the specific physics associated to
the symmetry of a
$d$-wave superconducting order parameter, because 
transport through the  ferromagnetic
electrodes has a strong directional dependence. 
The orientation of the electrodes can be used to probe the 
symmetry of the order parameter. 

The line shape of the conductance spectra is V-like 
which is a fingerprint of $d$-wave systems. 
This has already been tested in experiments \cite{covington}.

Moreover we used a theoretical description that is valid
not only in the tunnel regime but also for large
interface transparencies. We have found 
a ZEP in the conductances ${\cal G}_{a,b}$, ${\cal G}_{a,a}$
in the tunnel regime in the two cases of parallel and
antiparallel spin orientations, for the $\beta=\pi/4$ orientation.

\section{conclusions}
Using a Keldysh formalism we have shown that 
in the ferromagnet/$d$-wave superconductor/ferromagnet junction, 
transport
is due to crossed Andreev reflection and elastic cotunneling
and is mediated by ZES that are formed at the interface due to the
sign change of the order parameter.

We have used the local 
surface Green's function given by Eq. (\ref{dwaveg}) and we find no 
particular relation between the conductances in the parallel 
and antiparallel alignments. In the $s$-wave case and for 
extended contacts it is possible to show that the average current 
due to crossed Andreev reflection in the antiparallel alignment 
is equal to the average current due to elastic cotunneling in the 
parallel alignment \cite{melin2,falci}. Discussing 
multichannel effects
for $d$-wave superconductors is left as an important
open question.

\end{document}